\documentstyle[12pt,epsfig]{article}             
\newcommand{\beq}{\begin{equation}}             
\newcommand{\eeq}{\end{equation}}               
\newcommand{\bqry}{\begin{eqnarray}}            
\newcommand{\eqry}{\end{eqnarray}}              
\newcommand{\bqryn}{\begin{eqnarray*}}          
\newcommand{\eqryn}{\end{eqnarray*}}            
\newcommand{\preprint}[1]{\begin{table}[t]      
            \begin{flushright}                  
            \begin{large}{#1}\end{large}        
            \end{flushright}                    
            \end{table}}                        
\newcommand{\PD}[2]                             
    {\frac{\partial^{#2}}{\partial #1^{#2}}}    
\begin{document} 
\preprint{LA-UR-99-6772} 
\title{Analysis of Dislocation Mechanism for Melting \\ of Elements: 
Pressure Dependence} 
\author{\\ Leonid Burakovsky\thanks{E-mail: BURAKOV@T5.LANL.GOV}, \
Dean L. Preston\thanks{E-mail: DEAN@LANL.GOV}, \
and Richard R. Silbar\thanks{E-mail: SILBAR@WHISTLESOFT.COM. Also at 
WhistleSoft, Inc., 
Los Alamos, NM 87544, USA
}
 \\  \\ 
Los Alamos National Laboratory \\ Los Alamos, NM 87545, USA }
\date{ }
\maketitle
\begin{abstract}
In the framework of melting as a dislocation-mediated phase transition we 
derive an equation for the pressure dependence of the melting temperatures 
of the elements valid up to pressures of order their ambient bulk moduli. 
Melting curves are calculated for Al, Mg, Ni, Pb, the iron group (Fe, Ru, Os), 
the chromium group (Cr, Mo, W), the copper group (Cu, Ag, Au), noble gases 
(Ne, Ar, Kr, Xe, Rn), and six actinides (Am, Cm, Np, Pa, Th, U). These 
calculated melting curves are in good agreement with existing data. We also 
discuss the apparent equivalence of our melting relation and the Lindemann 
criterion, and the lack of the rigorous proof of their equivalence. We show 
that the would-be mathematical equivalence of both formulas must manifest 
itself in a new relation between the Gr\"{u}neisen constant, bulk and 
shear moduli, and the pressure derivative of the shear modulus. 
\end{abstract}
\bigskip 
\centerline{{\it Key words:} melting, string, dislocation, melting curve, 
equation of state, high} 
\hspace*{3cm} pressure, elements, actinides 

\centerline{PACS: 62.50.+p, 64.10.+h, 64.70.Dv, 64.90.+b, 74.62.Fj, 77.84.Bw, 
91.60.Gf} 
\bigskip

\section{Introduction} 

The idea that a proliferation of dislocations is associated with melting dates 
back to Mott \cite{Mott}. The very first theory of dislocation-mediated 
melting \cite{MO} was a success, inasmuch as it predicted a first-order 
transition, as a consequence of incorporating the mutual screening of 
dislocations, in agreement with observations. Molecular dynamics \cite{MD} 
and Monte Carlo \cite{MC} calculations have more recently provided further 
evidence for the notion that dislocations drive the melting transition in 
three dimensions. There is also some experimental evidence that line defects 
are present in solids near melting \cite{Craw}. 

In refs.\ \cite{prev1,prev2} we formulated a dislocation theory of melting in 
which dislocations near melt were modeled as non-interacting strings on a 
lattice. The possible configurations of a dislocation were taken to be closed 
random walks. Screening of long-range strain fields by other dislocations in 
a dense ensemble 
results in a $-\rho \ln \rho $ dependence of the free energy on the 
dislocation density, $\rho ,$ and thus a first-order transition. We obtained 
the following relation between the melting temperature $T_m,$ the shear 
modulus, $G,$ the Wigner-Seitz volume, $v_{WS},$ the coordination number, 
$z,$ and the critical density of dislocations, $\rho (T_m)$ (in units where 
$k_B=1):$ 
\beq 
T_m=\frac{\kappa \lambda Gv_{WS}}{8\pi \ln (z-1)}\;\!\ln \left( \frac{\alpha 
^2}{4b^2\rho (T_m)}\right) . 
\eeq 
Here $b$ is the length of the shortest perfect-dislocation Burgers vector, 
$\kappa $ is 1 for a screw dislocation and $(1-\nu )^{-1}\approx 3/2$ for an 
edge dislocation $(\nu $ is the Poisson ratio), $\lambda \equiv b^3/v_{WS}$ 
and $\alpha ,$ which accounts for non-linear effects in the dislocation core, 
has a value of 2.9 \cite{prev2}. Experimental data on 51 elements show that 
\beq 
\frac{Gv_{WS}}{4\pi T_m\ln (z-1)}=1.01\pm 0.17 
\eeq 
at zero pressure \cite{prev1}. Eqs.\ (1) and (2) imply that the critical 
dislocation density at zero pressure is 
\beq
\rho (T_m)=(0.61\pm 0.20)\;\!b^{-2}. 
\eeq
This value is in good agreement with the critical density
\beq
\rho (T_m)=(0.66\pm 0.11)\;\!b^{-2}, 
\eeq 
obtained by applying our relation for the latent heat of fusion \cite{prev2}, 
\beq 
L_m=\frac{1}{\lambda}\;\!b^2\rho (T_m)R\;\!T_m\ln (z-1), 
\eeq 
to data on latent heats for 75 elements. Hence, $b^2\rho (T_m)$ is 
approximately constant across the Periodic Table with the numerical value 
\beq
b^2\rho (T_m)=0.64\pm 0.14, 
\eeq
which is the uncertainty-weighted average of Eqs.\ (3) and (4). 

In this paper we investigate the validity of our melting relation, Eq.\ (1), 
up to pressures of order 100 GPa, by comparing to experimental melting curves, 
i.e., melting temperatures versus pressure, $p.$ This comparison requires 
$v_{WS}(p),$ or its equivalent, the pressure dependence of the compression, 
$\eta \equiv V_0/V.$ We obtain $\eta (p)$ from the bulk modulus, $B(p),$ which 
is extrapolated to high pressure using only its value and first pressure 
derivative at ambient conditions, viz., room temperature and zero pressure. 
The shear modulus $G(p)$ is similarly extrapolated to high pressure. Pressure 
derivatives of $G$ and $B$ are typically $O(1),$ so the extrapolation of the 
bulk modulus is expected to break down at pressures of order the ambient bulk 
modulus. The parameter $\kappa $ in Eq.\ (1), which depends on the Poisson 
ratio, varies by only a few percent between $p=0$ and 100 GPa (we discuss this 
in more detail in Section 2). Since the accuracy of our melting relation at 
zero pressure is 17\%, we take $\kappa $ to be a constant. We also make the 
necessary but reasonable assumption that $b^2\rho (T_m)$ is also a 
pressure-independent constant. With this assumption, we find that our melting 
relation agrees well with experimental melting curves up to pressures $\approx 
B,$ and, in fact, our extrapolation of $T_m$ is often in good agreement with 
data to pressures $\approx 2B.$ In addition to the good agreement with the 
existing melting curve data, we also predict the high-pressure melting 
curves of Ag, Au, Cr, Cu, Mo, Os, Ru, W, and several actinides. 
 
\section{Melting curve equation}

We now consider the pressure dependences of the factors appearing in our 
melting relation, Eq.\ (1). The parameter $\lambda $ is constant by its 
definition, $\alpha $ is also assumed to be a constant, and $\kappa $ may 
be taken as constant provided that the Poisson ratio $\nu $ has a very 
weak pressure dependence. In fact, for an isotropic medium \cite{Gschn} 
\beq
\nu =\frac{1}{2}\;\!\frac{3B-2G}{3B+G}. 
\eeq
Although both $G$ and $B$ vary with pressure, the ratio in Eq.\ (7) varies 
only weakly. Consider, for example, Cu, for which $\nu \approx 0.34$ 
at $p=0.$ At $p=100$ GPa, we calculate the values of $G$ and $B$ with the 
help of Eqs.\ (13) and (14) below, with their pressure derivatives taken from 
ref.\ \cite{GS}, and find $\nu \approx 0.38.$ Therefore, in this case the 
corresponding values of $1/\kappa \approx 1-\nu /2$ \cite{prev2} 
are 0.83 and 0.81, respectively, so that the variation in the average value of 
$1/\kappa $ is $\approx 2$\%. Thus, the pressure dependence of $1/\kappa $ can 
be safely neglected. (There exists an upper bound on the change in the value 
of $1/\kappa $ with pressure. In the ultra-high-pressure limit, $p\propto 
\eta ^{5/3},$ in agreement with 
the theory of the free electron gas (Fermi gas). Therefore, $B\equiv -Vdp/dV=
\eta \;\!dp/d\eta \propto \eta ^{5/3}\gg G\propto \eta ^{4/3},$ and hence $\nu 
\rightarrow 1/2,$ in view of Eq.\ (7) (see also \cite{Kop}). Thus, in contrast 
to the Poisson ratio which changes by $\approx 50$\%, $1/\kappa \approx 1-\nu 
/2$ changes by $\approx 10$\%: $5/6\rightarrow 3/4.)$ 

We assume further that the mean interdislocation spacing at the melting point, 
$R\approx 1/\sqrt{\rho (T_m)},$ scales with $b,$ independent of pressure, and 
hence $b^2\rho (T_m)$ is a pressure-independent constant (with a numerical 
value of $0.64\pm 0.14,$ in view of Eq.\ (6)). It then follows from Eq.\ (1) 
that, provided the coordination number does not change with pressure 
(i.e., the element either remains in the same crystalline phase, or changes 
phase without changing the coordination number, e.g., a face-centered cubic 
structure $\leftrightarrow $ a hexagonal close-packed structure), the melting 
relation is given by 
\beq
\frac{G(p,T_m(p))v_{WS}(p,T_m(p))}{T_m(p)}={\rm const.} 
\eeq

The dependence of $v_{WS}$ on pressure and temperature is just the equation 
of state of the metal. Let us first focus on its temperature dependence. The 
fixed-pressure ratio of Wigner-Seitz volumes at $T_m$ and $T=0$ is equal to 
$1+\beta T_m,$ where $\beta $ is the volume expansivity. At $p=0,$ $\beta $
is typically of order $10^{-5}$ K$^{-1},$ and melting temperatures are at most 
about 4000 K, so $v_{WS}$ changes by only a few percent between $T=0$ and 
$T_m.$ Assuming that $\beta $ does not increase appreciably with compression, 
we can use room-temperature values for $v_{WS}.$ 

In contrast to $v_{WS},$ the dependence of $G$ on $T$ is not necessarily weak. 
Its $T$-dependence involves two characteristic temperatures, namely the Debye 
temperature, $T_D,$ and the melting temperature. $G$ is always monotonically 
decreasing with $T,$ and is nonlinear for $T\stackrel{<}{\sim }T_D$ and 
linear from $T_D$ to $T_m.$ However, there are no experimental data, no 
computer calculations, and no theoretical guidance that tells us how the 
temperature dependence of $G$ varies with pressure. In particular, how does 
the (negative) slope of the linear region vary with $p$? At this point we have 
no choice but to conjecture. We assume that $G(p,T_m(p))/G(p,0)$ is a slowly 
varying function of $p,$ so it can be considered constant up to moderate 
compressions, say, 20\% to 30\%. Thus, $G(p,T_m)$ is replaced by $G(p,0)$ 
in Eq.\ (8). In addition, data on the $p=0$ temperature dependence of shear 
moduli \cite{SW} clearly show that $G(p,300)\approx G(p,0),$ and therefore, 
we use the room temperature value of the shear modulus in our melting relation.

Subsequently, the explicit dependence of $G$ and $v_{WS}$ on $T$ will be 
dropped. It will be understood that $G$ and $v_{WS}$ are at room temperature. 
Our melting relation now reads 
\beq 
\frac{G(p)v_{WS}(p)}{T_m(p)}={\rm const.} 
\eeq

Differentiating Eq.\ (9) with respect to $p,$ one finds 
\beq 
\frac{1}{T_m}\;\!\frac{dT_m}{dp}=\frac{1}{G}\;\!\frac{dG}{dp}-\frac{1}{B}, 
\eeq
where we have used the definition of the bulk modulus, 
\beq 
B(p)\equiv -V\frac{dp}{dV}=-v_{WS}\frac{dp}{dv_{WS}}. 
\eeq
Thus, upon integration, Eq.\ (10) gives 
\beq
\frac{T_m(p)}{T_m(0)}=\frac{G(p)}{G(0)}\;\!\exp \;\!\left\{ -\int _0^p
\frac{dp'}{B(p')}\right\} . 
\eeq
To proceed further, we have to specify $G(p)$ and $B(p).$ 

\subsection{The shear modulus $G$ at finite pressure}

For the shear modulus at all pressures, we use the relation \cite{GS2} 
\beq
G=G_0+G'_0\;\!\frac{p}{\eta ^{1/3}}, 
\eeq
where $G_0'\equiv (dG/dp)_0.$ 
The subscript 0 refers to ambient conditions: $T\simeq 300$ K and $p=0.$ 

This equation satisfies the requirement that 
$G\propto \eta ^{4/3}$ as $\eta \rightarrow \infty ,$ since $p\propto \eta ^{
5/3}.$ With the values of $G'_0$ for 32 elements tested in ref.\ \cite{GS2} 
Eq.\ (13) gives nearly the right value for the proportionality constant 
between $G$ and $\eta ^{4/3}$ at high compressions. Eq.\ (13) works 
well for a diverse selection of engineering metals covering many different 
crystal structures and nearly all groups of the Periodic Table \cite{GS2}. 

\subsection{Compression and the bulk modulus $B$ at finite pressure} 

Expanding the bulk modulus around $p=0$ we have 
\beq 
B(p)=B_0+B_0'p+\frac{1}{2}\;\!B_0''p^2+\ldots , 
\eeq 
where $B_0$ and $B_0'\equiv (dB/dp)_0,$ $B_0''\equiv (d^2B/dp^2)_0,\ldots $ 
can be extracted from equation of state data. Values of $B_0''$ are known 
for a few elements only (their determination is highly uncertain and involves 
an error of order 100\% \cite{FI}), and besides, $B_0''$ first appears in the 
$(p/B_0)^3$ term in the power series expansion of $\eta :$ 
$$\eta =\exp \;\!\left\{ \int _0^p\frac{dp'}{B(p')}\right\}=\left[ \frac{2B_0+
(B_0'+\sqrt{B_0'^2-2B_0B_0''})\;\!p}{2B_0+(B_0'-\sqrt{B_0'^2-2B_0B_0''})\;\!p}
\right] ^{1/\sqrt{B_0'^2-2B_0B_0''}}$$ 
\beq 
=1+\left( \frac{p}{B_0}\right) -\frac{B_0'-1}{2}\left( \frac{p}{B_0}\right) ^2
+\frac{(B_0'-1)(2B_0'-1)-B_0B_0''}{6}\left( \frac{p}{B_0}\right) ^3+\ldots \;. 
\eeq 
Since only $B_0$ and $B_0'$ are generally known (for almost all the elements, 
see ref.\ \cite{GS}), we restrict ourselves instead to the first two terms in 
Eq.\ (14). Then the compression simplifies to 
\beq
\eta =\left( 1+\frac{B_0'}{B_0}\;\!p\right) ^{1/B'_0}\!. 
\eeq 
Eqs.\ (15) and (16) are two different approximations to 
the Murnaghan equation of state \cite{Murn,HS}. 

It then follows from Eqs.\ (12), (13) and (16) that the equation of the 
melting curve is 
\beq
T_m(p)=T_m(0)\left( 1+\frac{B'_0}{B_0}\;\!p\right) ^{-1/B'_0}\left[ 1+\frac{
G'_0}{G_0}\;\!p\left( 1+\frac{B'_0}{B_0}\;\!p\right) 
^{-1/3B'_0}\right] .
\eeq
As discussed in Section 4, this equation is only valid for pressures 
$p\stackrel{<}{\sim }2B.$ 

It follows from (17) that for $p\ll B_0$ 
\beq 
T_m(p)=T_m(0)\left[ 1+\left( \frac{B_0G_0'}{G_0}-1\right) \left( \frac{p}{B_0}
\right) -\left( \frac{4}{3}\;\!\frac{B_0G_0'}{G_0}-\frac{B_0'+1}{2}\right) 
\left( \frac{p}{B_0}\right) ^2+\ldots \right] . 
\eeq 
For the vast majority of the elements, $B_0'>5/3$ and $B'$ approaches 5/3 in 
the limit of large compressions. (In this limit $p\propto \eta ^{5/3},$ and 
therefore $B\equiv -Vdp/dV=\eta \;\!dp/d\eta =5p/3,$ i.e., $B'=5/3.)$ In fact, 
the average value of $B_0'$ for the 65 elements analyzed in \cite{GS}, except 
for Ce for which $B_0'<0,$ is $4.30\pm 1.40.$ Hence, if 
\beq 
\frac{G_0'}{G_0}>\frac{3}{8}\;\!\frac{B_0'+1}{B_0}, 
\eeq 
it follows from $B_0'>5/3$ that also $G_0'/G_0>1/B_0,$ i.e., Eq.\ (18) is of 
the form $T_m(p)=T_m(0)(1+ap-bp^2+\ldots ),$ $a,b>0,$ and describes melting 
curves for which melting temperatures increase with pressure \cite{Young}. 
If, however, 
\beq 
\frac{G_0'}{G_0}<\frac{1}{B_0} 
\eeq 
and $B_0'>5/3,$ then also  $G_0'/G_0<3/8\;\!(B_0'+1)/B_0,$ i.e., Eq.\ (18) is 
of the form $T_m(p)=T_m(0)(1-ap+bp^2-\ldots ),$ $a,b>0,$ and describes melting 
curves for which melting temperatures initially decrease with pressure 
\cite{Young}. For Si, for example, with the data from ref.\ \cite{GS} we find 
$G_0'/G_0<1/B_0$ and $B_0'=4.19,$ in agreement with the negative initial slope 
of the experimental melting curve. Eqs.\ (19) and (20) plus $B_0'>5/3$ 
should be considered our criteria for the two types of melting curves 
discussed above. 

\section{Melting curves: comparison with data} 

In this section we compare our melting curve, Eq.\ (17), to some experimental
melting curves, and predict a number of melting curves that can be compared 
with experiment in the not-so-distant future. 

We have found 5 elements for which melting curves have been measured to higher 
pressures, $p\sim O(100$ GPa): Al, Fe, Ni, Pb and U. We compare experimental 
data for these elements with our curves in Figs.\ 1-5. For Al, we also show 
the best fit to data in the form of the Simon equation, $T_m(p)=T_m(0)(1+ap)^
b$ \cite{Al}. For Fe, the experimental data are from ref.\ \cite{Fe}, and from 
ref.\ \cite{Ni} for Ni. For Pb, we combine the high-pressure data of ref.\ 
\cite{Pb} with the low-pressure data of ref.\ \cite{Pb2} as corrected in ref.\
\cite{Pb3}. For U, the high-pressure data of ref.\ \cite{U} are combined with 
the low-pressure data of ref.\ \cite{U2}. 

As claimed in ref.\ \cite{Al}, the Simon equation may not be the best 
functional form for a fit to data. In fact, the initial slope provided by this 
equation for the Al melting curve is 80 K/GPa, in contrast to 59 and 65 K/GPa 
from the two previous low-pressure measurements \cite{Al}. This accounts for 
the difference between the two curves in Fig.\ 1. 

In Fig.\ 4, in addition to Fe, we also plot melting curves for Ru and Os, 
elements in the same column of the Periodic Table. Those curves should be 
considered predictions for these metals. 

In Fig.\ 5, in addition to U for which there are high-pressure data, we also 
plot melting curves for the 5 actinides Am, Cm, Np, Pa and Th. For Np, we also 
show the low-pressure data of ref.\ \cite{Np}. We do not show the low-pressure 
data of ref.\ \cite{Am} for Am since they would overlay the low-pressure U 
data. We have checked that our melting curve is in agreement with the 
low-pressure Am data. For Cm, the values of $B_0$ and $B_0'$ are taken from 
ref.\ \cite{Cm}, and the value of $G_0$ is that estimated in ref.\ 
\cite{prev1}. For Pa, the values of $B_0$ and $G_0$ come from ref.\ 
\cite{Gschn}. We estimate the values of $G_0'$ for Cm and Pa from Th and U, 
their neighbors in the same row in the Periodic Table. Our earlier $G_0'$ 
estimates for Am and Np lead to $\gamma =1.05$ and 1.09, respectively, in 
Eq.\ (24), which implies that such estimates are reliable. The values of 
$B_0'$ for Np and Pa are also estimated by interpolating between Am, Cm, Th 
and U. (We note that this estimation of $B_0'$ is justified by the pronounced 
periodic behavior of $B_0'$ in $Z$ \cite{Stein}.) The value of $B_0'$ for Am 
is taken from \cite{Am}. We emphasize that the predicted melting curves assume 
constancy of coordination number along them. In the case of Am, e.g., there is 
still disagreement over the correct sequence of phases and their transition 
pressures \cite{Young}, so this assumption may well be incorrect. For Th, 
however, it is claimed that there is a transition from a face-centered cubic 
structure to a body-centered tetragonal structure that changes coordination 
number \cite{Vohra}. This transition occurs in the pressure range of $70-100$ 
GPa \cite{Vohra}, and thus our predictions for the Th melting curve up to 
75 GPa should be quite reliable. 

Although we can account for a decrease in melting temperature with pressure 
in our theoretical framework (Eqs. (17),(20)), we do not consider such cases 
here, among which there are Pu and Ce. It has been established  \cite{Th} 
that for Th, which is in the same column as Ce, $\triangle V>0,$ and 
therefore, in view of Eq.\ (23), its melting temperature increases with 
pressure. 

%
\begin{center}
\vspace{2cm} 
\epsfig{file=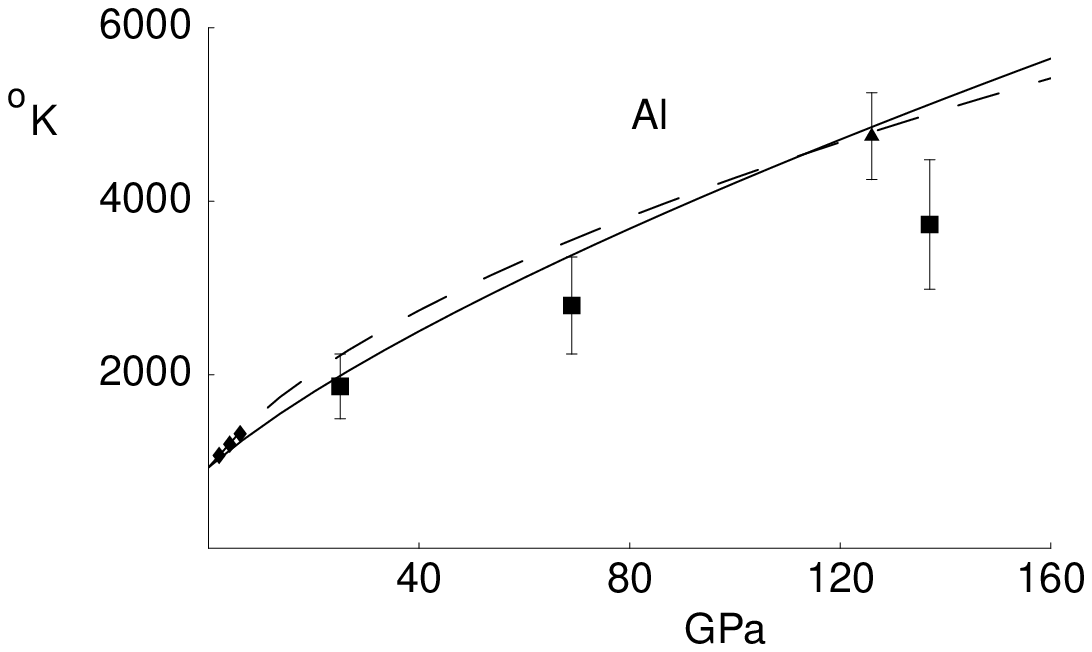,width=15cm,angle=0}
\end{center}
Fig.\ 1. Melting curve for Al. The dashed line is the Simon-fit to the data 
of ref.\ \cite{Al}, which are not shown explicitly. The diamonds are the 
low-pressure data from ref.\ \cite{LW}. The triangle is the shock-melting 
point at 125 GPa from ref.\ \cite{Al3}. The boxes are the points at 25, 69 
and 137 GPa calculated in ref.\ \cite{Urlin} from shock-melting data. 
They are assigned 20\% error bars \cite{Urlin}. 
 \\

\begin{center}
\vspace{-0.3cm} 
\epsfig{file=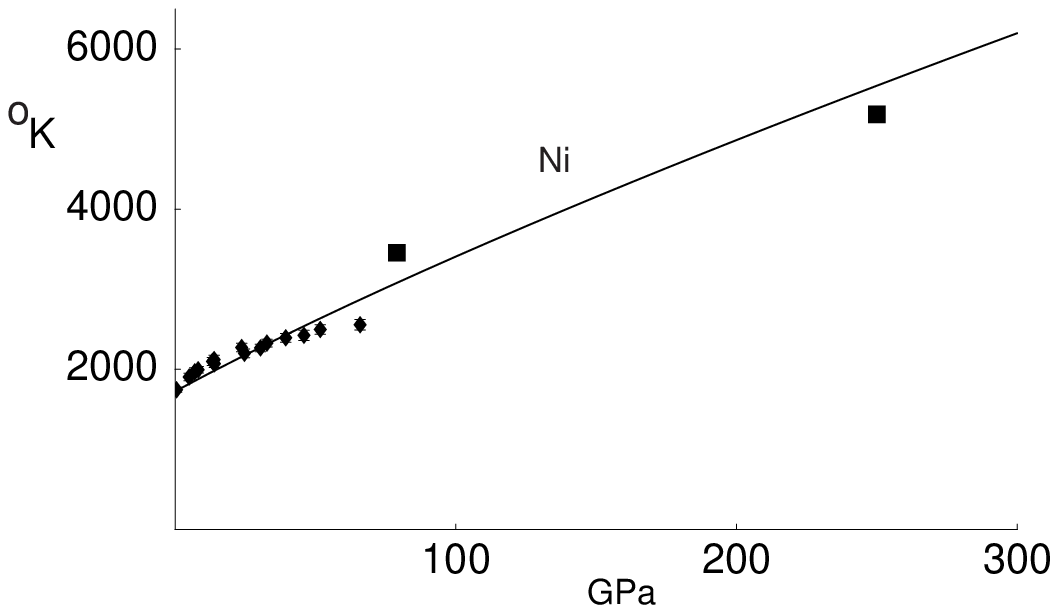,width=15cm,angle=0}
\end{center} 
Fig.\ 2. Melting curve for Ni. The diamonds (with small error bars) are the 
data of ref.\ \cite{Ni}. The boxes are the points at 79 and 250 GPa calculated 
in ref.\ \cite{Urlin} from shock-melting data. The corresponding error bars 
are not quoted in ref.\ \cite{Urlin}. 
 \\

\begin{center}
\epsfig{file=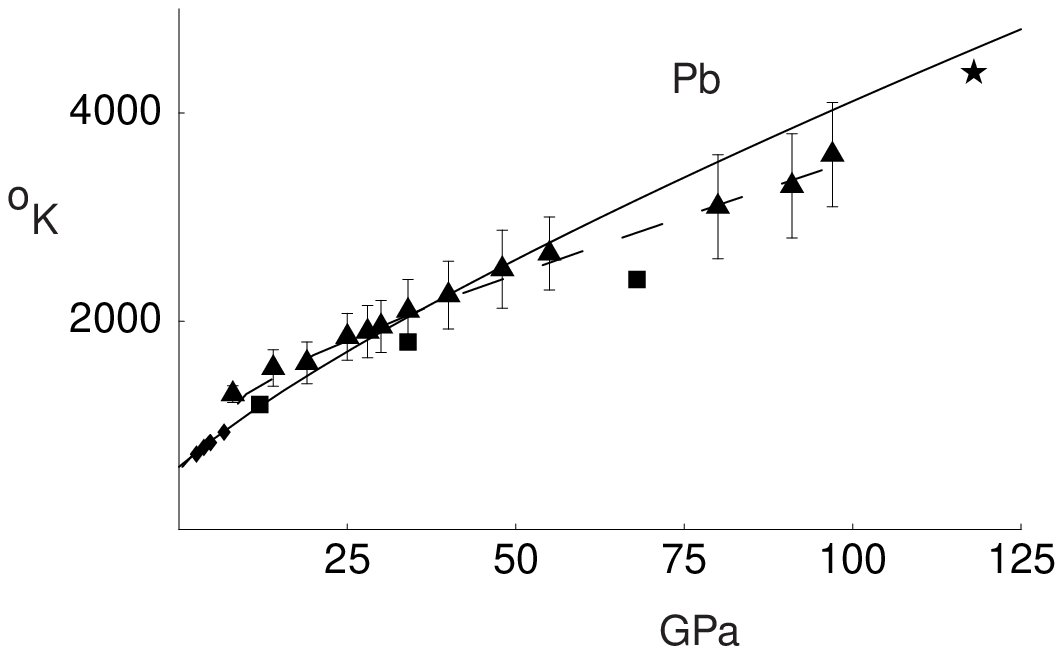,width=15cm,angle=0}
\end{center}
Fig.\ 3. Melting curve for Pb. The diamonds are the low-pressure data of 
ref.\ \cite{Pb2} corrected as in ref.\ \cite{Pb3}. The triangles are the data 
from ref.\ \cite{Pb}, and the dashed line is a best fit \cite{Pb} to the data.
The boxes are the points at 12, 34 and 68 GPa calculated in ref.\ \cite{Urlin} 
from shock-melting data. The corresponding error bars are not quoted 
in ref.\ \cite{Urlin}. The star is the point at 118 GPa calculated in ref.\ 
\cite{God} from shock-melting data. 
 \\

\begin{center}
\vspace{2cm} 
\epsfig{file=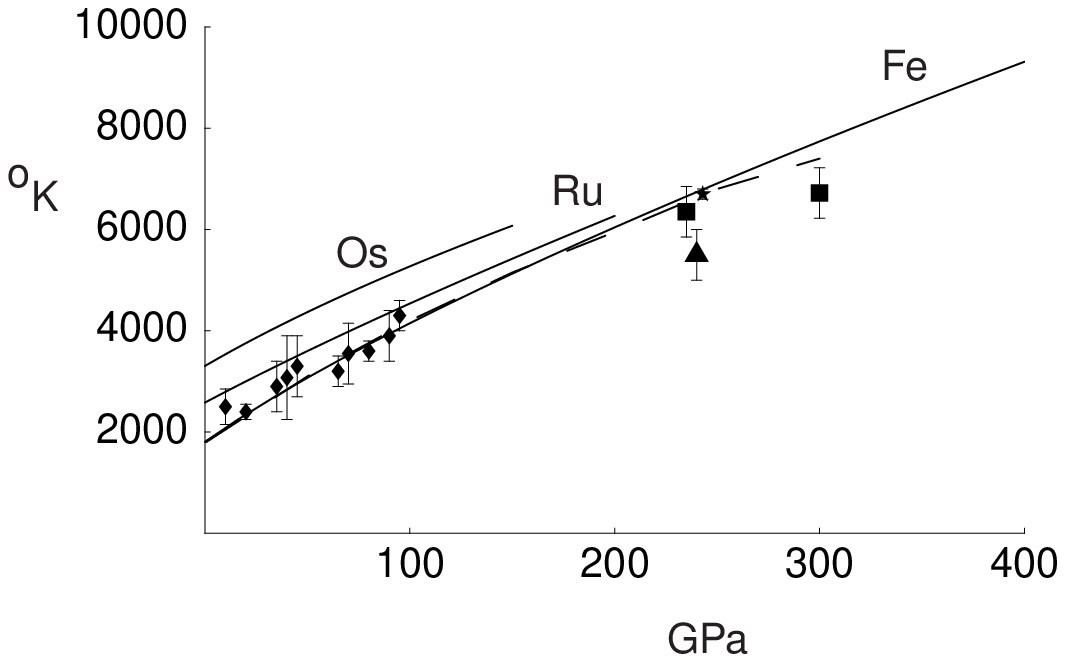,width=15cm,angle=0}
\end{center}
Fig.\ 4. Melting curves for the elements of the iron group (Fe, Ru, Os). The 
diamonds are the data of ref.\ \cite{Fe}, and the dashed line is a best fit 
\cite{Fe} to the data. The boxes are the shock-melting points at 235 GPa and 
300 GPa \cite{Fe2}. The triangle is the shock-melting point at 240 GPa 
\cite{Fe3}. The star is the shock-melting point at 243 GPa \cite{Fe4}. 
 \\

\begin{center}
\vspace{0.5cm} 
\epsfig{file=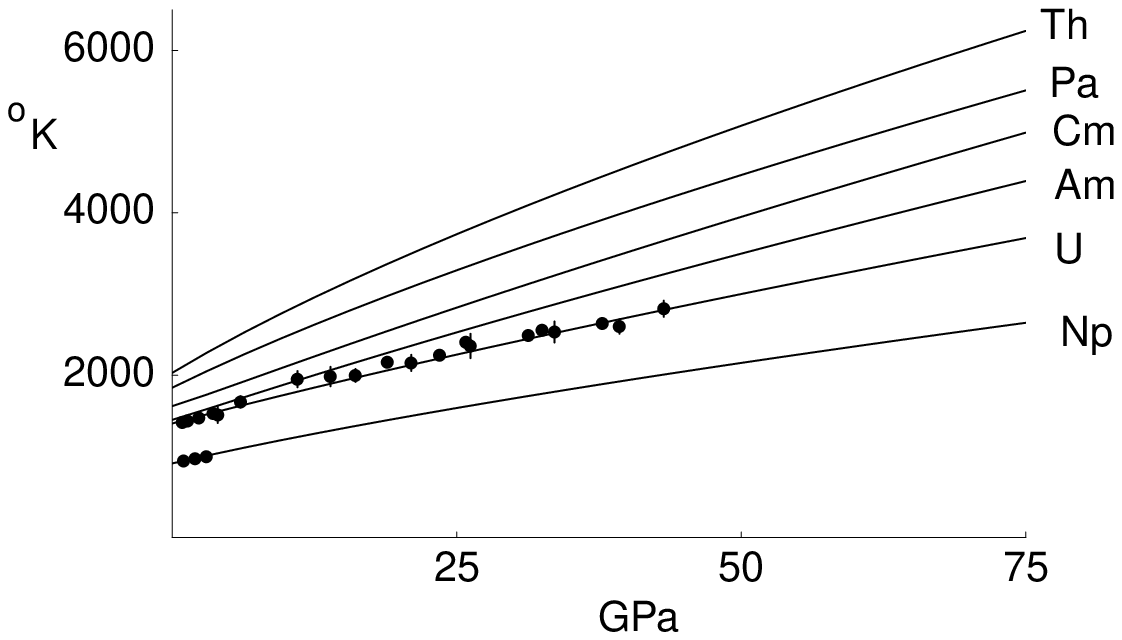,width=15cm,angle=0}
\end{center}
Fig.\ 5. Melting curves for the actinides Am, Cm, Np, Pa, Th and U. The data 
for U are from ref.\ \cite{U}, and for Np they come from ref.\ \cite{Np}. 
 \\

In Fig.\ 6 we plot the low-pressure data of ref.\ \cite{W} for W, and our 
melting curves for W, Mo and Cr. The initial slope of our melting curve of 
Mo, 26 K/GPa, is consistent with that predicted in ref.\ \cite{Mo}: 
$(34\pm 6)$ K/GPa. The same melting curve gives $T_m\simeq 9650$ K at $p=390$ 
GPa, in good agreement with the shock-melting temperature $\sim 10000$ K at 
the same pressure, found in ref.\ \cite{Mo2}. 

%
\begin{center}
\vspace{2cm} 
\epsfig{file=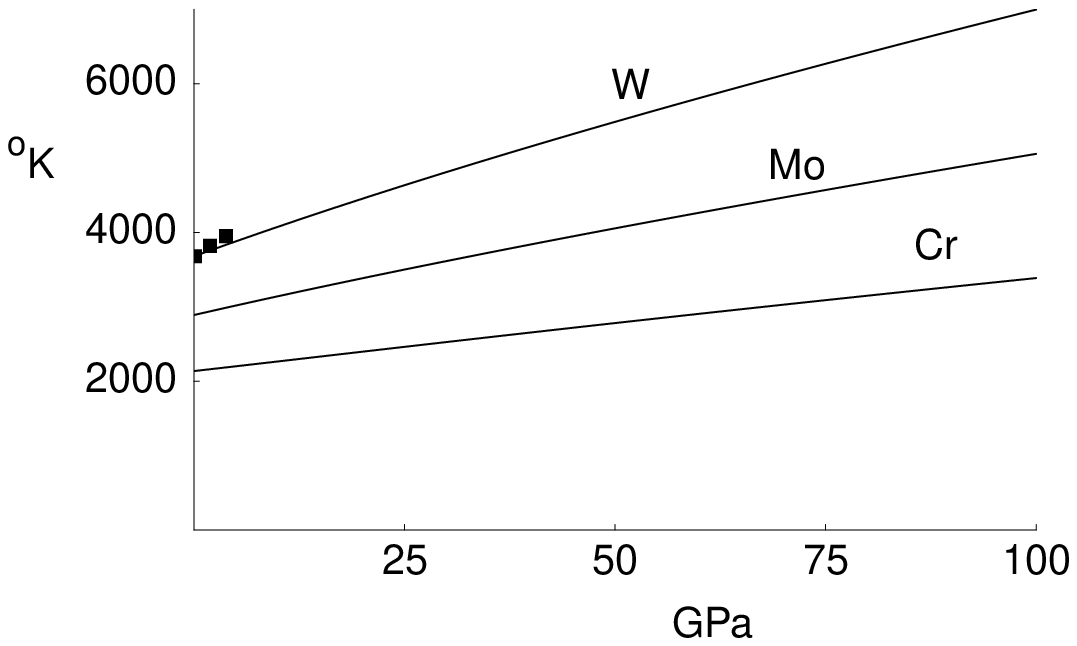,width=15cm,angle=0}
\end{center}
Fig.\ 6. Melting curves for the elements of the chromium group (Cr, Mo, W). 
The data for W are from ref.\ \cite{W}. 
 \\

In Fig.\ 7 we compare the low-pressure data of ref.\ \cite{CuAgAu} for Cu, Ag 
and Au with our corresponding melting curves. Although the initial slopes of 
these curves are somewhat less than those of the data (the corresponding 
values of $\gamma $ in Fig.\ 1 are $\simeq 0.8),$ they are in good agreement 
with the best extrapolation of data to higher pressures made in ref.\ 
\cite{CuAgAu}, and with the calculation of ref.\ \cite{Urlin} in the case of 
Cu. 

%
\begin{center}
\vspace{2cm} 
\epsfig{file=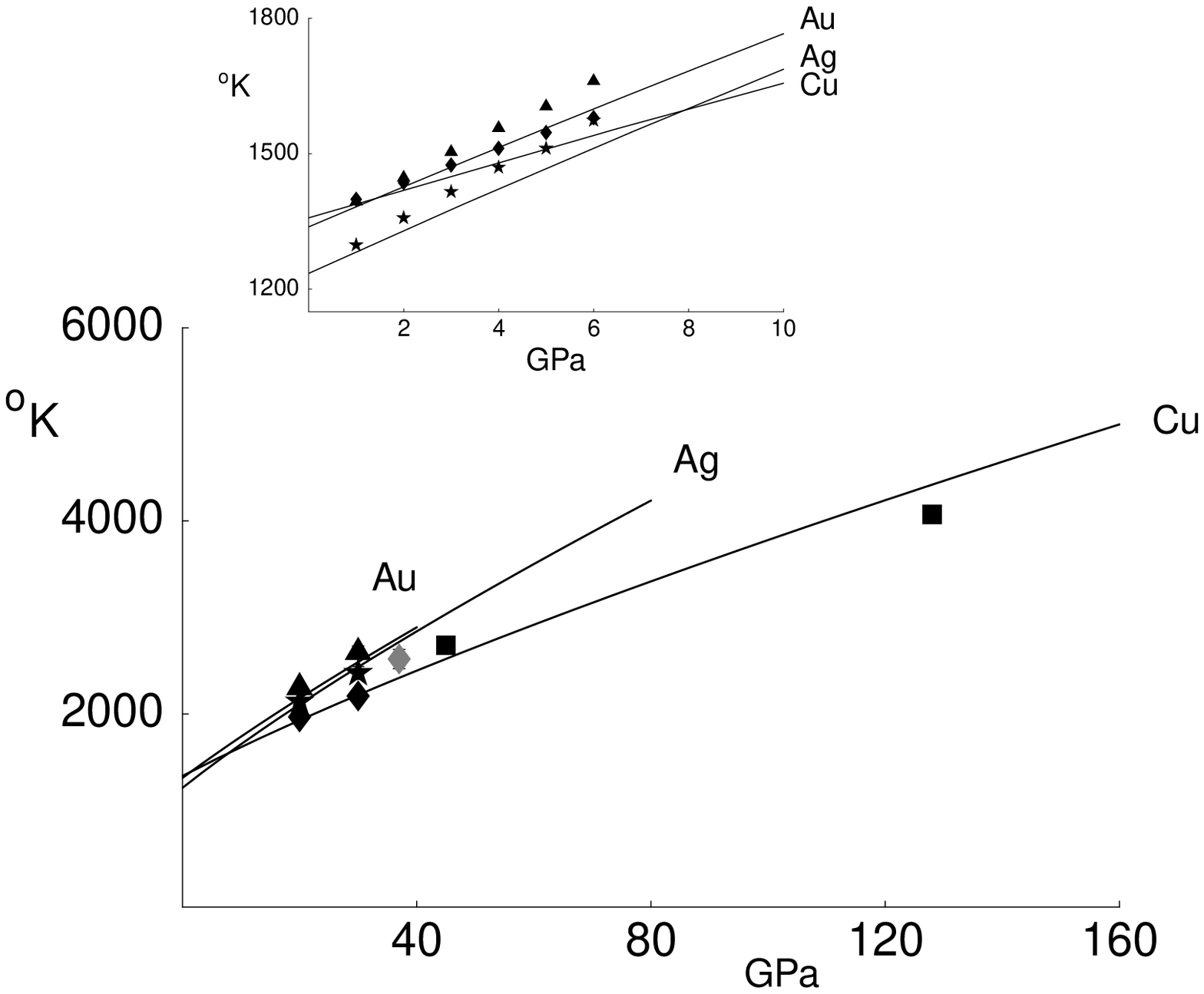,width=15cm,angle=0}
\end{center}
Fig.\ 7. Melting curves for the elements of the copper group (Cu, Ag, 
Au). The diamonds, the stars, and the triangles are the low-pressure data of 
ref.\ \cite{CuAgAu} in the inset, and the best extrapolations of these data 
to 20 and 30 GPa in the main plot for Cu, Ag and Au, respectively. The boxes 
are the points at 45 and 128 GPa calculated in ref.\ \cite{Urlin} from 
shock-melting data for Cu. The corresponding error bars are not quoted 
\cite{Urlin}. The gray diamond is the shock-melting point for Cu 
at 37 GPa \cite{Cu2}.
 \\ 

In Fig.\ 8 we compare the low-pressure data on the noble gases to our 
corresponding melting curves. The unknown values of $G'_0$ for Ne, Ar, Kr 
and Xe are calculated with the help of Eq.\ (25) below using the measured 
values of $B_0,$ $G_0$ and $\gamma _0$ (the Gr\"{u}neisen constant) 
\cite{prev2}. The values of $B_0'$ for Ne, Ar, Kr and Xe are taken from 
\cite{B'}. In the case of Rn, for which $B_0,$ $B_0',$ $G_0$ and $G_0'$ have 
not been measured, we first calculate $G_0$ using the (approximate) relation 
$GV/T_m={\rm const}$ for the noble-gas group, where $V=v_{WS}N_A$ is the 
molar volume. This relation follows from Eq.\ (1) provided that $\kappa ,$ 
$\lambda ,$ $\alpha $ and $z$ do not vary within this group. The value of the 
constant is determined by using the corresponding Ne, Ar, Kr and Xe data in 
this relation. We then calculate $B_0$ using Eq.\ (7) with the value of the 
Poisson ratio for Rn determined by extrapolating from the corresponding values 
for Ne, Ar, Kr and Xe. Finally, we determine both $B_0'$ and $G_0'$ by again 
extrapolating the Ne, Ar, Kr and Xe data. 

%
\begin{center}
\vspace{2cm} 
\epsfig{file=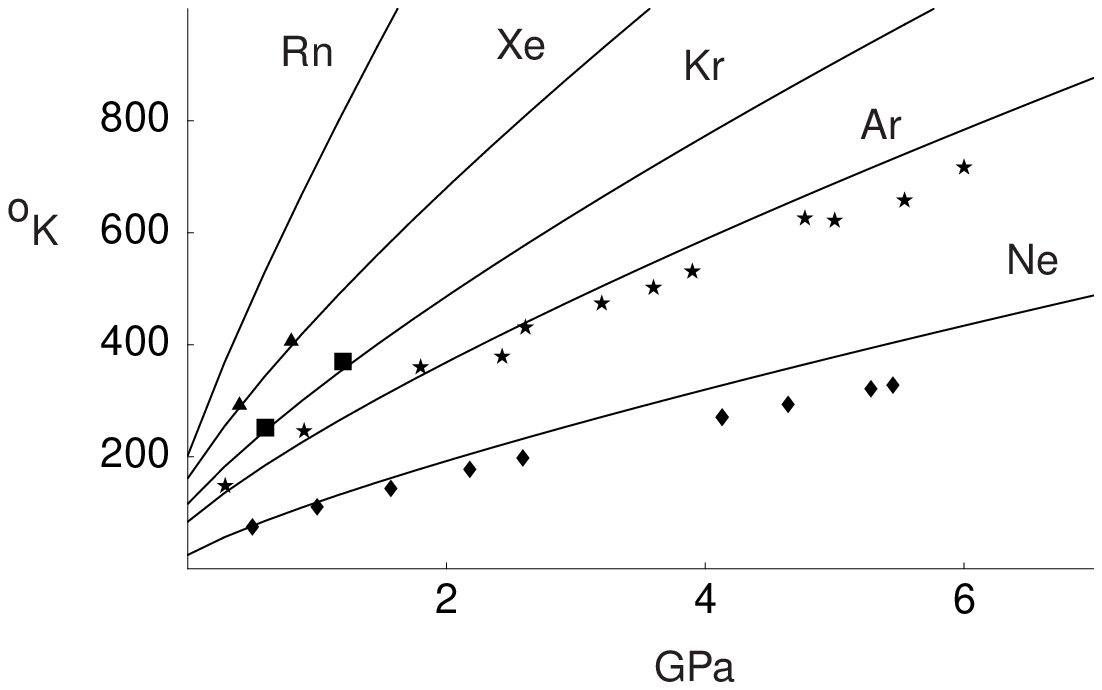,width=15cm,angle=0}
\end{center}
Fig.\ 8. Melting curves for the noble gases Ne, Ar, Kr, Xe and Rn. 
The diamonds are the data of ref.\ \cite{Ne}. The stars are the data of 
ref.\ \cite{Ar}. The boxes and triangles come from ref.\ \cite{KrXe}. 
 \\ 

Finally, in Fig.\ 9 we show the experimental data and our theoretical 
melting curve for Mg. 

%
\begin{center}
\vspace{-0.8cm} 
\epsfig{file=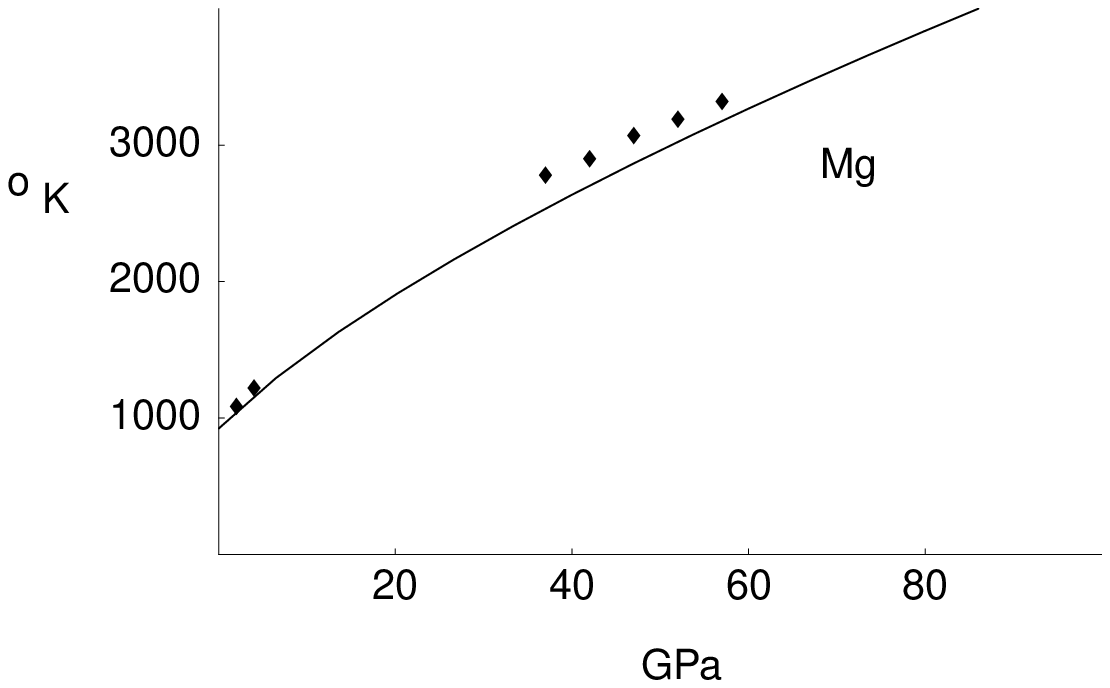,width=15cm,angle=0}
\end{center}
Fig.\ 9. Melting curve for Mg. The low-pressure data are from ref.\ 
\cite{Mg_}. The high-pressure data are the shock-melting points of ref.\ 
\cite{Mg^}.
 \\ 

The 24 melting curves considered above constitute convincing evidence for the 
validity of our formula for melting temperature as a function of pressure, 
Eq.\ (17).  

\section{The range of validity of the new melting curve equation} 

In deriving our melting curve, Eq.\ (17), we have used both Eq.\ (13) for 
the pressure dependence of the shear modulus and the Murnaghan equation of 
state, Eq.\ (16). Since Eq.\ (13) has the correct zero-pressure limit (its 
Taylor series expansion in $p$ at $p=0$ is $G=G_0+G_0'p-(G_0'/3B_0)p^2+\ldots 
)$ and is claimed to have the correct ultra-high-pressure limit \cite{GS2}, 
we assume that this equation is valid over the entire pressure range. In any 
event, we do not have data to either confirm or invalidate this assumption. It 
then follows that the range of validity of Eq.\ (17) depends crucially on the 
range of validity of the Murnaghan equation of state, Eq.\ (16). 
 
The Murnaghan equation of state was examined in ref.\ \cite{HS}, together 
with a number of different equations of state, by comparing with the 
theoretical results calculated by the augmented-plane-wave method and the 
quantum-mechanical model proposed by Kalitkin and Kuz'mina \cite{KK} from 
low to ultra-high pressures. It was shown that the Murnaghan equation 
is in good agreement with the theoretical results up to $V/V_0\simeq 0.7,$ 
i.e., up to compressions $\simeq 1.4-1.5.$ Since for the vast majority of the 
elements $B_0'\approx 5$ \cite{GS}, 
we conclude, on the basis of Eq.\ (16), that the Murnaghan equation, and 
consequently, our equation for melting curve, Eq.\ (17), is valid up to 
pressures $p\approx 2B_0.$ The melting curves for Al and Pb in Figs.\ 1 and 3, 
respectively, and for Ne and Ar in Fig.\ 8 show that in some cases Eq.\ (17) 
is good to pressures even greater than $2B_0.$ 

For reliable predictions of melting curves to much higher pressures, 
$p\stackrel{>}{\sim }1$ TPa, one has to use a better equation of state than 
Murnaghan's. Hama and Suito \cite{HS} claim that the Vinet equation of state 
\cite{Vinet} is consistent with first-principles theoretical calculations to 
compressions $\eta \sim 5.$ Reference \cite{CGH} also finds the Vinet equation 
of state to be most accurate among various suggested equations of state. 
In fact, we have calculated that the melting temperatures for Fe and Pb at 
pressures $p\sim 50\;\!B_0$ as given by Eq.\ (17) are about two times higher 
than those given by a relation that derives from Eqs.\ (12), (13), and the 
Vinet equation of state. 

Another possible source of disagreement between the new melting curve, Eq.\ 
(17), and data may be inaccurate values of elastic constants and their pressure
derivatives in some cases. The Murnaghan equation and its frequently used 
partner -- the Birch equation \cite{Birch} -- are derived from the 
second-order Taylor series expansion of the bulk modulus [as in Eq.\ (14)] or 
the elastic strain energy with respect to pressure or strain, respectively. 
Thus their validities are, in principle, restricted to a narrow range of 
compression. Extending this range would entail the inclusion of higher-order 
terms. This could explain why the values of $B_0,$ and especially those of 
$B_0'$ and $B_0'',$ obtained from experiments which cover different ranges of 
compression by using a fitting method, are usually different. In many cases 
these differences between different experiments are small and can be safely 
neglected. In some cases, however, they are large, and so their use for 
predicting physical observables, such as melting temperature, is dubious. 
For example, in the case of Ni, we have used the value $B_0'=6.20$ given 
in \cite{GS}. Reference \cite{Ger}, however, quotes $B_0'\simeq 30$ (!). 
Similarly, for Mo we have used $B_0'=4.4$ of ref.\ \cite{GS}, while ref.\ 
\cite{Ger} gives $B_0'\simeq 20.$ (We note that the use of the values $B_0'=
6.20$ for Ni and 4.4 for Mo is justified in view of the recent compilations 
of experimental data on $B_0'$ \cite{RMR}.) Although the numerical value of 
$B_0'$ does not matter at low $p,$ since it first appears in the $(p/B_0)^2$ 
term, in view of Eq.\ (15), it would strongly affect the predicted melting 
curve at pressures $p\sim O(B_0).$ 
 
There are also inconsistencies in the values of $G_0$ quoted in the 
literature. For example, for Pb we use the value $G_0=8.6$ GPa from ref.\ 
\cite{GS}, whereas ref.\ \cite{Gschn} quotes $G=5.5$ GPa. (Our own calculation 
\cite{prev1}, based on the values of the elastic constants $c_{11},$ $c_{12}$ 
and $c_{44},$ shows that 8.6 GPa is preferred over 5.5 GPa.) Likewise the 
values of $G_0$ for K and Na from ref.\ \cite{GS} are 0.9 and 1.98 GPa, 
whereas ref.\ \cite{Gschn} quotes 1.3 and 3.5 GPa, respectively. 

\section{Relation of dislocation-based melting relation to the Lindemann 
criterion} 

The well-known Lindemann melting rule is based on the assumption that all 
elemental solids melt when the atomic vibrational amplitude is a fixed 
pressure-independent fraction of the interatomic distance. As shown by 
Lindemann \cite{Lin}, this implies the invariance of the Lindemann number 
\beq
\theta _D\left( \frac{M}{T_m}\right) ^{1/2}V^{1/3}=L
\eeq 
along the melting curve. Here $\theta _D$ is the density-dependent Debye 
temperature, $V$ is the molar volume, and $M$ is the molar mass. It is found 
that $L\approx 150$ \cite{Gschn}. 

There are compelling reasons to suppose that our dislocation-based melting 
relation is somehow equivalent to the Lindemann criterion. First of all, Eq.\ 
(21) gives melting curves that are typically very close to those predicted by 
our dislocation-based melting relation. (For example, the melting curve for Mg 
in Fig.\ 6.3 of ref.\ \cite{Young} is {\it very} similar to our curve in Fig.\ 
9.) Furthermore, the Lindemann number, which is proportional to the ratio of 
atomic vibrational amplitude to the lattice constant at melt, is analogous to 
$b^2\rho (T_m),$ since both are presumed constant along the melting curve. 
In the dislocation-based approach, melting is associated with a critical 
configuration of dislocations, and for any such configuration there is a 
corresponding mean displacement of atoms from their equilibrium positions. 
Hence $L$ and $b^2\rho (T_m)$ are clearly related, and therefore the 
left-hand sides of Eqs.\ (9) and (21) are related as well. 

The mathematical equivalence of our melting relation and the Lindemann 
criterion would be established if it  could be determined that the left-hand 
side of Eq.\ (21) is a fixed fraction of the left-hand side of Eq.\ (9). A 
search of the literature has turned up two results which show that $L^2$ is 
{\it approximately} proportional to $Gv_{WS}/T_m.$ For a Debye solid the 
relation is \cite{Klein} $L^2=f(\nu (p,T))Gv_{WS}/T_m,$ where $f$ is a 
complicated function of $\nu .$ Thus the two melting relations are not 
rigorously equivalent. 

A second connection between the two melting formulas is provided by the 
following approximation for the Gr\"{u}neisen constant \cite{GS}, 
\beq 
\gamma (p)=\frac{2}{3}\;\!\gamma _S(p)+\frac{1}{3}\;\!\gamma _L(p), 
\eeq 
where
\beq 
\gamma _S(p)=\frac{G'(p)}{2}\;\!\frac{B^T(p)}{G(p)}-\frac{1}{6}, 
\eeq 
\beq 
\gamma _L(p)=\frac{1}{2}\;\!\frac{B^T(p)}{B^S(p)+\frac{4}{3}G(p)}\;\!\frac{
d(B^S(p)+\frac{4}{3}G(p))}{dp}-\frac{1}{6}, 
\eeq 
are the contributions of the shear (transverse) and longitudinal acoustic 
modes. Here $B^T$ is the isothermal bulk modulus, which is equivalent to 
$B$ that we are using in this paper. Eqs.\ (22)-(24) follow from the two 
assumptions that (i) the only appreciable contribution to the heat capacity 
of a crystal arises from lattice vibrations, and (ii) averaging over all modes 
is equivalent to averaging only over the low-frequency acoustic modes. (I.e., 
the contribution of the optical modes is equal to that of the acoustic modes.)
If in addition it is assumed that $B^S(p),$ the isentropic bulk modulus, is 
proportional to $G(p),$ then 
\beq 
\gamma _S(p)=\gamma _L(p)=\gamma (p)=\frac{G'(p)}{2}\;\!\frac{B(p)}{G(p)}-
\frac{1}{6}. 
\eeq 
However, there is no basis for this assumption, i.e., $\gamma _S(p)\neq \gamma 
_L(p)$ is to be expected. For example, in the ultra-high pressure limit, 
$B^S(p)\sim B^T(p)=5p/3$ and $G(p)\sim p^{4/5},$ quite different dependencies. 

Integration of Eq.\ (25), using $B(p)=-dp/d\ln V(p)$ and $\gamma (p)=-d\ln 
\theta _D(p)/d\ln V(p),$ gives 
\beq 
\frac{\theta _D^2(p)V^{2/3}(p)/T_m(p)}{G(p)V(p)/T_m(p)}={\rm const,} 
\eeq
that is, $L^2\propto Gv_{WS}/T_m.$ We emphasize that this proportionality 
is founded on a number of uncontrolled approximations. 
 
Equation (25), which would ensure a rigorous mathematical equivalence of 
the defect and mechanical (Lindemann's) approaches to melting, does {\it not} 
follow from first principles. 
This means that the defect and mechanical approaches to melting are basically 
{\it different.} Moreover, since the mechanical approach does not have a solid 
thermodynamic basis, it cannot, for example, predict the latent heat of 
fusion. In contrast, the defect approach 
predicts the latent heat of fusion, Eq.\ (9), which is in good 
agreement with data for three-quarters of the Periodic Table \cite{prev2}. 

Finally, we wish to make the following comments on Eq.\ (25). We did not check 
extensively its validity at zero pressure, 
since that would go beyond the scope of this paper. We do, however, 
have some evidence that Eq.\ (25) is rather well satisfied: 
with the data from ref.\ \cite{GS}, we calculate from the above relation 
$\gamma _0=2.25$ vs.\ measured 2.40 for Ag, 3.08 vs.\ 2.99 for Au, 1.66 vs.\ 
1.78 for Fe, 1.28 vs.\ 1.29 for K, and 1.18 vs.\ 1.19 for Na. We have actually 
used Eq.\ (25) in Section 3 to calculate $G_0'$ for noble gases in order to 
get their melting curves via Eq.\ (17) and to compare with experiment. Good 
agreement between the calculated and experimental curves is another hint 
on the approximate validity of this formula. Also, Eq.\ (25) has the correct 
ultra-high-pressure limit in which $\gamma \rightarrow 1/2$ \cite{Kop}, 
since in this limit $G\sim p^{4/5}$ and $B=5p/3.$ 

\section{Concluding remarks} 

We have extended the framework of melting as a string-mediated phase 
transition to non-zero pressure and derived a new equation for the 
melting curve, Eq.\ (17). As discussed above, with accurate experimental 
values of all the parameters involved, this equation reproduces the 
existing experimental melting data, and predicts unknown melting curves to 
pressures $p\stackrel{<}{\sim }2B_0.$ For higher pressures, a better equation 
of state than Murnaghan's should be used, e.g., the Vinet equation of state. 

We have addressed the apparent equivalence of defect and mechanical 
approaches to melting curve, and demonstrated that both approaches are 
basically different. We have shown that their would-be rigorous mathematical 
equivalence must manifest itself in a new relation, Eq.\ (25), 
which we have not tested in detail. 

To summarize, we have calculated melting curves for 24 elements: Al, Mg, Ni, 
Pb, the iron group (Fe, Ru, Os), the chromium group (Cr, Mo, W), the copper 
group (Cu, Ag, Au), the noble gases Ne, Ar, Kr, Xe and Rn, and the six 
actinides Am, Cm, Np, Pa, Th and U. These calculated melting curves are 
in good agreement with existing data. 
 
\section*{Acknowledgements} 

We wish to thank T. Goldman and A.Z. Patashinski for valuable discussions 
during the preparation of this work. One of us (L.B.) wishes to thank 
B.K. Godwal for useful correspondence.

\end{document}